# Dynamic Channel Allocation for QoS Provisioning in Visible Light Communication

Mostafa Zaman Chowdhury, Muhammad Shahin Uddin, and Yeong Min Jang
Department of Electronics Engineering, Kookmin University, Seoul, Korea

*Abstract*--**In visible light communication (VLC) diverse types of traffic are supported while the number of optical channels is limited. In this paper we propose a dynamic channel reservation scheme for higher priority calls that does not reduce the channel utilization. The number of reserved channels for each traffic class is calculated using real time observation of the call arrival rates of each traffic classes. The numerical results show that the proposed scheme is able to reduce the call blocking probability of the higher priority user within a reasonable range without sacrificing channel utilization.**

## I. INTRODUCTION

VLC system is one type of optical wireless communication system in which visible light is used as a transmission medium. The use of VLC is harmless to humans, provides high security and high data rates, license free frequency band and does not cause the malfunction of aircraft equipment or medical instruments [1]. In VLC system each channel has its own color according to the draft VLC specification frequency band plan in [2]. If multi-class of traffic such as voice, video, data, etc. are aggregated in the system, then each call of different traffic classes can use different color channel. In VLC system resource is allocated for each types of user based on the priority and instantaneous color channel condition for maintaining the Quality of Service (QoS). To provide QoS for the higher priority users in VLC system, the approach to give privileged for the higher priority users is needed to reduce their call blocking rate. The fixed guard channels always reduce the resource utilization. Thus, we need a dynamic guard channel scheme that can reduce the call blocking rate of higher priority calls within a reasonable range without sacrificing huge resources. The dynamic channel allocation scheme in this paper reserves a non-fixed numbers of channels to prioritize higher priority calls. The instantaneous number of reserved channels for a call of a particular class of traffic is calculated using the call arriving rates of different classes of traffic and total allocated channel. The priority scheme using the dynamic channel reservation is applicable for the higher traffic condition only.

## II. PROPOSED RESOURCE ALLOCATION SCHEME

The fixed guard channel to prioritize any call always reduces the resource utilization [3]. For the VLC networks, an efficient mechanism is needed that can ensure very low call blocking rate for the higher priority users without sacrificing resource utilization. In this section, we propose a non-fixed guard channel scheme. The dedicated channels for a particular class of traffic are varied according to the variation of call arriving rates of all the traffic classes. Let traffic class *1, 2, 3, ..., m, ..., M* represent the highest to lowest priority traffic classes respectively. $\lambda_T$ and $\lambda_m$ represents total call arriving rate and call arriving rate of traffic class *m* respectively. So

$$\lambda_T = \lambda_1 + \lambda_2 + \cdots + \lambda_m + \cdots + \lambda_M \quad (1)$$

Suppose *N* and *C* represent total number of channels in the system and fixed number of minimum common channels allocation for all the traffic classes respectively. The numbers of reserved channels for each of the traffic classes are calculated on the two following basic conditions.

*Light traffic load condition:* This condition is defined as

$$\lambda_T < \frac{N}{\Gamma} \quad (2)$$

where $\Gamma$ is a threshold value (around 0.9~0.95 times of the average channel holding time) to decide the traffic condition.

For the light traffic load condition, there is no priority scheme that maximizes the channel utilization. For this case all the traffic classes can use whole of *N* channels.

*High traffic load condition:* This condition is defined as

$$\lambda_T \geq \frac{N}{\Gamma} \quad (3)$$

The number of reserved channels for only the traffic class *1* i.e. the highest priority traffic class is

$$X_1(t) = \frac{\lambda_1}{\lambda_T}(N - C) \quad (4)$$

The number of channels available for the traffic class *2* is

$$N_2(t) = N - X_1(t) \quad (5)$$

The number of reserved channels for only the traffic classes of *1, 2, ..., and* $(m-1)$ is

$$X_{m-1}(t) = \frac{\lambda_{m-1}}{\lambda_T}(N - C) \quad (6)$$

If we assume $X_0(t) = 0$, then total number of available channels for the traffic class *m* (m-th priority class traffic) is

$$N_m(t) = N - [X_1(t) + X_2(t) + \cdots + X_{m-1}(t)] \quad (7)$$

Equations (4) to (7) show that the channel allocations for different traffic classes are not fixed. Depending on $\lambda_T$, *N*, and *C*, the number of channel allocation is varied. At a particular time, the instantaneous number of maximum channels $N_m(t)$ for the traffic class *m* can be calculated as:

Let $\Delta t_m$ is the time interval of the last two call arrivals of traffic class *m*. Then $\lambda_1 = \frac{1}{\Delta t_1}, \lambda_2 = \frac{1}{\Delta t_2}, \ldots, \lambda_M = \frac{1}{\Delta t_M}$

From (4) to (7), we find that

$$N_m(t) = N - \frac{\frac{1}{\Delta t_1}+\frac{1}{\Delta t_2}+\cdots+\frac{1}{\Delta t_{m-1}}}{\frac{1}{\Delta t_1}+\frac{1}{\Delta t_2}+\cdots+\frac{1}{\Delta t_m}+\cdots+\frac{1}{\Delta t_M}}(N - C) \quad (8)$$

Here the value of $\Delta t_m$ is always varied and small value of

This work was supported by the IT R&D program of MKE/KEIT. [10035362, Development of Home Network Technology based on LED-ID]

$\Delta t_m$ refers higher call arrival rate of traffic class $m$. Thus the number of reserved channel for the traffic class $m$ is high if $m$ represents higher priority traffic class. Hence, the number of allocated channels for a traffic class is dynamically varied depending on the call arrival rates, $N$, and $C$. Fig. 1 shows the system model for the proposed dynamic channel allocation scheme in high traffic condition. The state transition rate diagram for the proposed scheme is shown in Fig. 2. This state transition rate diagram is valid for the high traffic condition only. $1/\mu$ is the average channel holding time in Fig. 2. Three classes of traffic i.e $M = 3$ is considered for the analysis.

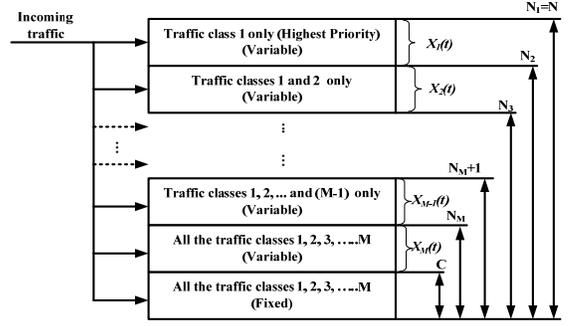

Fig. 1. System model for the proposed scheme.

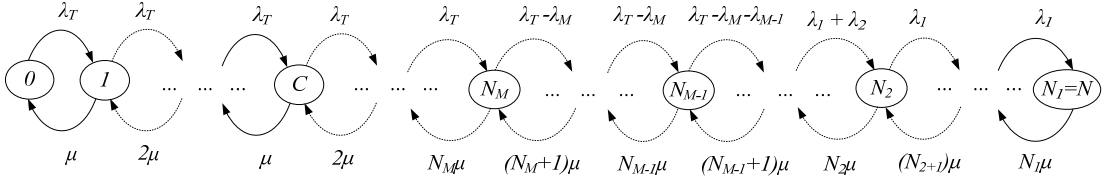

Fig. 2. State transition rate diagram for the proposed scheme

Suppose $P_i$ represents the probability that the system is in state $i$. The call blocking probability [4] for traffic class 1 is

$$B_1 = P_N = \frac{\lambda_x^{N_3} \lambda_y^{N_2-N_3} \lambda_1^{N-N_2}}{\mu^N N!} P_0 \quad (9)$$

Call blocking probability for traffic class 2 is

$$B_2 = \sum_{i=N_2}^{N} P_i = \sum_{i=N_2}^{N} \frac{\lambda_x^{N_3} \lambda_y^{N_2-N_3} \lambda_1^{i-N_2}}{\mu^N N!} P_0 \quad (10)$$

Call blocking probability for traffic class 3 is

$$B_3 = \sum_{i=N_3}^{N} P_i = B_2 + \sum_{i=N_3}^{N_2-1} \frac{\lambda_x^{N_3} \lambda_y^{i-N_3}}{\mu^i i!} P_0 \quad (11)$$

where

$$P_0 = \left[ \sum_{i=0}^{N_3} \frac{\lambda_x^i}{\mu^i i!} + \sum_{N_3+1}^{N_2} \frac{\lambda_x^{N_3} \lambda_y^{i-N_3}}{\mu^i i!} + \sum_{N_2+1}^{N} \frac{\lambda_x^{N_3} \lambda_y^{N_2-N_3} \lambda_1^{i-N_2}}{\mu^i i!} \right]^{-1}$$

and $\lambda_x = \lambda_1 + \lambda_2 + \lambda_3$, $\lambda_y = \lambda_1 + \lambda_2$.

### III. PERFORMANCE EVALUATION

In this section, we verified the performance of the proposed scheme. We consider voice (priority 1), video (priority 2), and data (priority 3) traffic classes for our analysis. $C$ is considered as 50% of $N$. Fig. 3 shows that the proposed scheme provides better call blocking probability for the higher priority users compared to the non-priority scheme. The proposed scheme also offers excellent channel utilization that is shown in Fig. 4

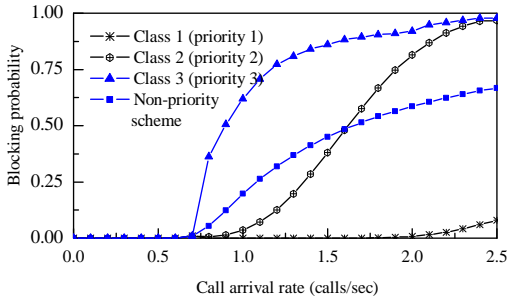

Fig. 3. Call blocking probability comparison

If the call arrival rate of any traffic class is increased or decreased, then the proposed scheme will also provide the similar performance as shown in the Fig. 3 and Fig. 4.

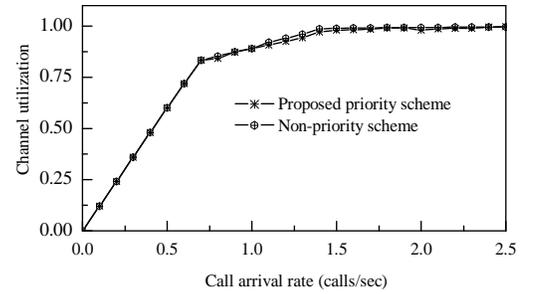

Fig. 4. Channel utilization comparison

### IV. CONCLUSIONS

The VLC system accommodates diverse of traffic classes. However, due to the limited number of channels, it is not possible to provide services for all the requested calls during high traffic condition. Hence, to provide QoS for the higher priority user, we need such a scheme which provides less blocking probability of the higher priority users without sacrificing channel utilization. Our proposed scheme provides better performance in terms of both the call blocking probability and channel utilization. Therefore, our proposed scheme is an excellent approach to apply in VLC system.